\documentclass[12pt]{article}
\begin{document}
\def\hatS{\hat s}
\def\hatT{\hat t}
\def\hatU{\hat u}
\def\cW{\cos \theta_W}
\def\sW{\sin \theta_W}
\def\warp{{\cal K} R_c \phi}
\def\warpp{{\pi {\cal K} R_c}}
\thispagestyle{empty}

\begin{flushright}
TIFR/TH/00-38 \\
IITK/PHY/2000/13 \\
{\bf hep-ph/0007354}
\end{flushright}

\bigskip
\bigskip

\begin{center}
{\LARGE\bf
Testing the Randall-Sundrum Model at a High Energy $\mathbf{e^- e^-}$ 
Collider} \\

\bigskip
\bigskip

{\sl Dilip Kumar Ghosh} \\

\bigskip
Department of Theoretical Physics, \\
Tata Institute of Fundamental Research, \\
Homi Bhabha Road, Mumbai 400 005, India. \\
E-mail: dghosh@theory.tifr.res.in \\
\bigskip
\bigskip
{\sl Sreerup Raychaudhuri}\footnote{On leave of absence from the 
Department of Physics, Indian Institute of Technology, 
Kanpur 208 016, India. E-mail: sreerup@iitk.ac.in} \\
\bigskip
Theory Division, CERN, \\ 
CH-1211 Geneva 23, Switzerland. \\
E-mail: sreerup@mail.cern.ch

\vskip 30pt

{\bf ABSTRACT}
\end{center}

\noindent
We study the process $e^- e^- \to e^- e^-$ at a high energy $e^- e^-$
collider including the effect of graviton exchanges in the warped
gravity model of Randall and Sundrum. Discovery limits for gravitons
are established and the effects of polarization are discussed.

\vskip 80pt

\begin{flushleft}
July 2000
\end{flushleft}

\vfill
\newpage 

\noindent
A great deal of recent interest centres around the physics possibilities
of a high energy linear collider with $e^\pm$ beams\cite{NLC}. Such a
machine can be run in $e^+e^-$ or $e^-e^-$ collision modes modes. Th
principal scattering processes at these are, respectively, Bhabha
scattering  $e^+e^- \to e^+e^-$ and M$\!\not\!{\rm o}$ller scattering $e^-
e^- \to e^- e^-$.

One of the useful features of M$\!\not\!{\rm o}$ller scattering $e^- e^-
\to e^- e^-$ at a high energy $e^- e^-$ collider is that it can receive
only a limited number of contributions from physics options\cite{Hsch}
which go beyond the Standard Model (SM). Among the interesting
beyond-Standard-Model (BSM) options are the exchange of multiple gravitons
in models with low-scale quantum gravity. The exchange of multiple
gravitons, in the $t$-channel as well as the $u$-channel, can affect the
process $e^- e^- \to e^- e^-$ in two ways:
\begin{itemize}
\item by changing (increasing or decreasing) the total cross-section from
the SM value --- this being the usual effect of BSM physics;
\item by changing the kinematic distributions of the final state electrons
--- this being the effect of exchanging particles with higher spin.
\end{itemize}

Two different scenarios of low-scale quantum gravity have attracted a
great deal of recent attention. In one of these, due to Arkani-Hamed,
Dimopoulos and Dvali (ADD)\cite{AkHd-Dmpl-Dvli}, one envisages a spacetime
with 4+$d$ dimensions, where the extra $d$ dimensions are compactified
with radii $R_c$ as large as a millimetre. In the ADD scenario, in four
dimensions there is a tower of massive Kaluza-Klein modes of the graviton,
whose masses are so densely-spaced (by as little as $10^{-13}$ GeV) as to
form a quasi-continuum. Though each graviton mode couples to electrons
with the feeble strength of Newtonian gravity, the collective effect of
all the gravitons contributes to interactions of almost electroweak
strength\cite{ADD-Revw}.  Effects of multiple exchange of gravitons in
M$\!\not\!{\rm o}$ller scattering, within the ADD scenario, have been
studied in Ref.\cite{Rizo}.

The other popular scenario of low-scale quantum gravity is that due to
Randall and Sundrum\cite{Rndl-Sndm}, who write a non-factorizable spacetime
metric
\begin{equation}
ds^2 = e^{-\warp} ~\eta_{\mu\nu} ~dx^{\mu} dx^{\nu} ~+~ R_c^2 ~d\phi^2
 \label{metric_RS}
\end{equation}
involving one extra dimension $\phi$ compactified with a radius $R_c$,
which is assumed to be marginally greater than the Planck length
$10^{-33}$ cm, and an extra mass scale ${\cal K}$, which is related to the
Planck scale $M^{(5)}_P$ in the five-dimensional bulk by ${\cal K}
\left[M^{(4)}_P\right]^2 \simeq \left[M^{(5)}_P\right]^3$.  Such a
`warped' geometry is motivated by compactifying the extra dimension on a
${\bf S}^1/{\bf Z}_2$ orbifold, with two $D$-branes at the orbifold fixed
points, {\em viz.}, one at $\phi = 0$ (`Planck brane' or `invisible
brane'), and one at $\phi = \pi$ (`TeV brane' or `visible brane').  The
interesting physical consequence of this geometry is that any mass scale
on the Planck brane gets scaled by the `warp factor' $e^{-\warpp}$ on the
TeV brane.  It now requires ${\cal K} R_c \simeq 12$ --- which is hardly
unnatural --- to obtain the hierarchy between the Planck scale and the
electroweak scale.  There still remains a minor problem: that of
stabilizing the radius $R_c$ (which is marginally smaller than the Planck
scale) against quantum fluctuations. A simple extension of the RS
construction involving an extra bulk scalar field has been
proposed~\cite{Gdbg-Wise-2} to stabilize $R_c$ and this predicts light
radion excitations with possible collider signatures~\cite{Mhnt-Rkst}.
Alternatively, supersymmetry on the branes can also act as a stabilizing
effect\cite{Bagr}. Models with SM gauge bosons and fermions in the bulk
have also been considered\cite{Ddsl-Hewt-Rizo-2}, but will not be
discussed in this work.

The mass spectrum and couplings of the graviton in the RS model have been
worked out, in Refs.~\cite{Gdbg-Wise,Ddsl-Hewt-Rizo}. We do not describe
the details of this calculation, but refer the reader to the original
literature.  It suffices here to note the following points.
\begin{enumerate}
\item There is a tower of massive Kaluza-Klein modes of the graviton, 
with masses
\begin{equation}
M_n = x_n {\cal K} e^{-\warpp}  \equiv  x_n m_o
   \label{mass}
\end{equation}
where $m_0 = {\cal K} e^{-\warpp}$ sets the scale of graviton masses and
is essentially a free parameter of the theory. The $x_n$ are the zeros of
the Bessel function $J_1(x)$ of order unity.
\item The massless Kaluza-Klein mode couples to matter with gravitational
strength; consequently its effects can be ignored for all practical
purposes.
\item Couplings of the massive Kaluza-Klein modes are gravitational,
scaled by the warp factor $e^{\warpp}$ and are consequently
of electroweak strength.
\end{enumerate}
Feynman rules (to the lowest order) for these modes have been worked out
in Refs.~\cite{THan-Lykn-Zhng}) and~\cite{Gudc-Ratz-Wels} in the context of
ADD-like scenarios. Each graviton couples to matter with strength $\kappa
= \sqrt{16\pi G_N}$.  All that we need to do to get the corresponding
Feynman rules in the RS model is to multiply the coupling constant $\kappa$
by the warp
factor $e^\warpp$ wherever necessary. It is convenient to write
\begin{equation}
\kappa e^{\warpp} = \sqrt{32\pi}~\frac{c_0}{m_0}
  \label{coupling}
\end{equation}
where $\kappa = \sqrt{16\pi G_N}$, using Eqn.~(\ref{mass}) and introducing
another undetermined parameter $c_0 \equiv {\cal K}/M^{(4)}_P$.  Thus
($c_0, m_0$)  may conveniently be taken as the free parameters of the
theory\footnote{ Though we differ from the exact choice of parameters in
Ref.~\cite{Ddsl-Hewt-Rizo}, a translation is easily made using the
formulae $c_0 = \frac{1}{8\pi} \left( {\cal K}/\overline{M}_P \right)$ and
$m_0 = \Lambda_\pi \left( {\cal K}/\overline{M}_P \right)$. It follows
that $c_0$ is roughly an order of magnitude less than ${\cal
K}/\overline{M}_P $ and $m_0$ can be one or two orders of magnitude
smaller than $\Lambda_\pi$.}.  Though $c_0$ and $m_0$ are not precisely
known, one can make estimates of their magnitude.  The RS construction
requires ${\cal K}$ to be at least an order of magnitude less than
$M^{(4)}_P$, which means that the range of interest for $c_0$ is about
0.01 to 0.1 (the lower value being determined by naturalness
considerations). $m_0$, which is of electroweak scale, may be considered
in the range of a few tens of GeV to a few TeV. Eq.~(\ref{mass}) tells us
that the first massive graviton lies at $M_1 = x_1 m_0 \simeq 3.83~m_0$.
Since no graviton resonances have been seen at LEP-2, running at energies
around 200 GeV, it is clear that we should expect $m_0 > 50$ GeV at least.

In this letter, we examine the effects of multiple graviton exchange in
M$\!\not\!{\rm o}$ller scattering in the RS scenario.  We focus on the
possibility of observing an excess in $e^- e^-$ events over the SM
prediction, and comment on possible refinements using the the kinematic
distributions of the final-state electrons. 
As earlier calculations\cite{Ddsl-Hewt-Rizo} have shown, in
the case when $c_0$ is large, the resonance structure in Bhabha scattering
is lost and there is not much difference, qualitatively speaking, between
Bhabha and M$\!\not\!{\rm o}$ller scattering in the RS model. In other
words, M$\!\not\!{\rm o}$ller scattering is as good a probe of this model
as Bhabha scattering in this case. It is on this option that our interest
is focussed.

The calculation of the Feynman amplitude involves, for the diagrams with
graviton exchange, a sum over graviton propagators of the form
\begin{equation}
\sum_n \frac{1}{t - M^2_n} 
~\equiv~ -~\frac{1}{m_0^2} ~\Lambda\left(\frac{\sqrt{-t}}{m_0}\right)
\end{equation}
and a similar sum with $t \leftrightarrow u$. Using the properties of the
zeros of Bessel functions, the function $\Lambda(x_t)$ can be written, to
a very good approximation, as\cite{PDas-RayC-Srkr}
\begin{equation}
\Lambda(x_t) = \frac{1}{\pi x_t} 
{\rm Im}~\psi\left(1.2331 + i~\frac{x_t}{\pi}\right) 
+ \frac{0.32586}{220.345 + 29.6898~x_t^2 + x_t^4}
\end{equation}
where $\psi(z)$ is the well-known digamma function. The variation of
$\Lambda(x_t)$ with $x_t$ is illustrated in Figure 1.  It is immediately
obvious that the effective coupling of the gravitons varies according to
the scattering angle, except in the case when $\sqrt{-t} \ll m_0$, i.e.
$x_t \to 0$. This is a feature quite different from that observed in the
related ADD model, where it is possible to take a limit in which a
similarly-defined $\lambda(x_t)$ is either constant or a slowly-varying
function. This is also a feature which can potentially change the angular
distribution of the final-state electrons.

There are  six Feynman diagrams corresponding to
M$\!\not\!\!{\rm o}$ller scattering
\begin{equation}
e^- (p_1, \lambda_1) ~+~ e^- (p_2, \lambda_2) \longrightarrow
e^- (p_3, \lambda_3) ~+~ e^- (p_4, \lambda_4)
\end{equation}
including the Standard Model as well as graviton-exchange diagrams.
Evaluation of these, using the Feynman rules for the RS model, and summing
over the final-state helicities $\lambda_3, \lambda_4$, is straightforward
and leads to a squared matrix element $|{\cal M}(\lambda_1,\lambda_2)|^2$,
whose explicit form is not given here in the interests of brevity.  If the
initial-state electrons have a left-handed longitudinal polarization $P$,
the differential cross-section is given by
\begin{eqnarray}
\frac{d\sigma}{dt} & = & \frac{1}{64\pi s^2}
\bigg[ (1 - P)^2 ~|{\cal M}(+,+)|^2 ~+~ (1 + P)^2 ~|{\cal M}(-,-)|^2 
\nonumber \\ & & \hskip 25pt
+ (1 - P^2) ~\left\{|{\cal M}(+,-)|^2 ~+~ |{\cal M}(-,+)|^2\right\} \bigg]
\end{eqnarray}

\begin{figure}[h]
\begin{center}
\vspace*{2.7in}
      \relax\noindent\hskip -3.0in\relax{\includegraphics{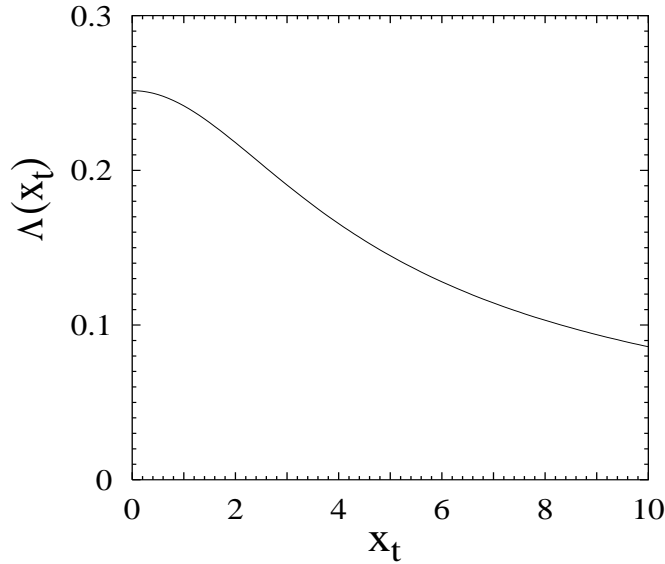}}
\end{center}
\caption{\footnotesize\sl
Illustrating the variation of the effective coupling $\Lambda$ of RS 
graviton towers to electrons as a function of $x_t = \sqrt{-t}/m_0$.}
\end{figure}

\noindent
assuming that both the beams are identically polarized. The importance of
the polarization factor $P$ is considerable, since it can be used, among
other things, to enhance or decrease the SM contribution to the
cross-section. In fact, polarization studies form an important part of the
physics program at a linear collider\cite{Plzn}.

In order to make a numerical estimate of the cross-section, we have
incorporated the calculated cross-section into a Monte Carlo event
generator, by means of which we calculate the cross-section for $e^- e^-
\to e^- e^-$ subject to the following kinematic cuts.
\begin{itemize}
\item The scattering angle of the electron(s) should not lie within $10^0$
of the beam pipe.
\item The transverse momentum of the electron(s) should not be less than 
10 GeV.
\end{itemize}
These `acceptance' cuts are more-or-less basic ones for any process at a
high-energy collider with electron and/or positrons. Though further
selection cuts will become appropriate when a more detailed analysis is
done, it suffices for our analysis, which is no more than a preliminary
study, to take the above cuts.  We then calculate the cross-section in the
SM and in the RS Model (including interference effects) for a fixed
polarization $P$ and given input parameters $c_0$ and $m_0$ of the RS
Model. Our results are given in Fig.~2.

In Fig. 2($a$), we present the total cross-section for the unpolarized
case $P~=~0$ as a function of machine energy for three different values of
the RS mass scale $m_0 = 150, 250$ and 500 GeV. The dashed line represents
the SM prediction and this exhibits the expected falling-off with machine
energy.  For large values of the graviton mass $m_0$, this behaviour is
preserved, since the graviton contribution is very small anyway. However,
when the graviton mass is smaller, the cross-sections show a marked
increase with energy, which reflects the well-known behaviour of gravity.
Obviously, at energies of 3--4 TeV, the gravitational contribution is huge
if the mass scale $m_0$ is small; however, a discernible difference exists
even when $m_0 = 500$ GeV. Thus, we can expect larger effects --- or,
conversely, stronger bounds --- on the RS Model as the machine energy
increases.

\begin{figure}[h]
\begin{center}
\vspace*{3.5in}
      \relax\noindent\hskip -3.5in\relax{\includegraphics{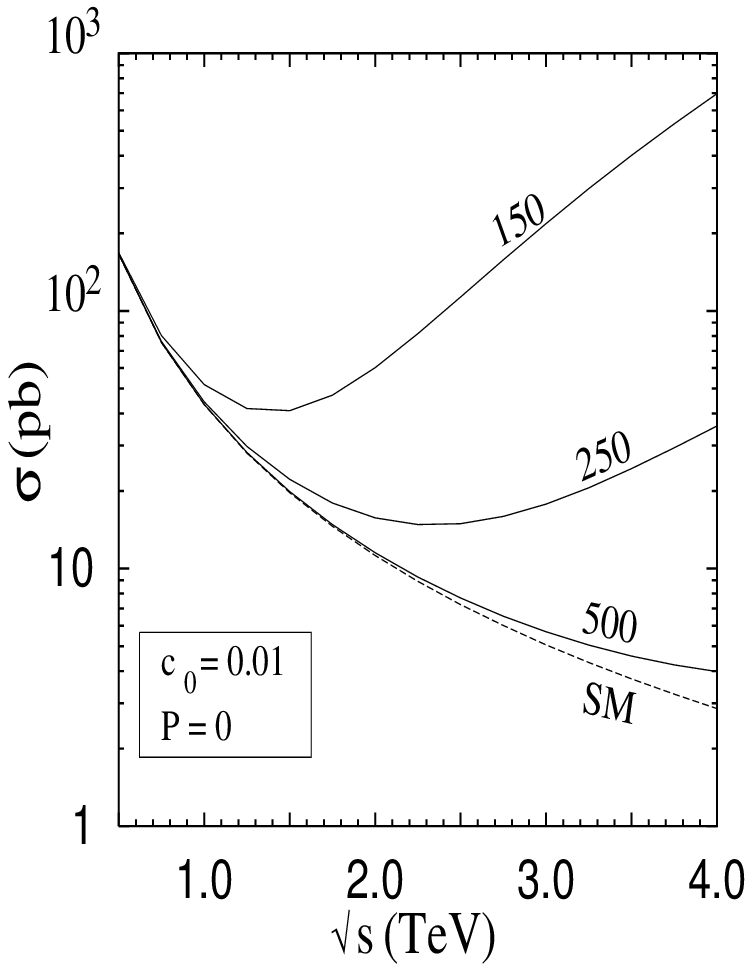}}
      \relax\noindent\hskip  3.4in\relax{\includegraphics{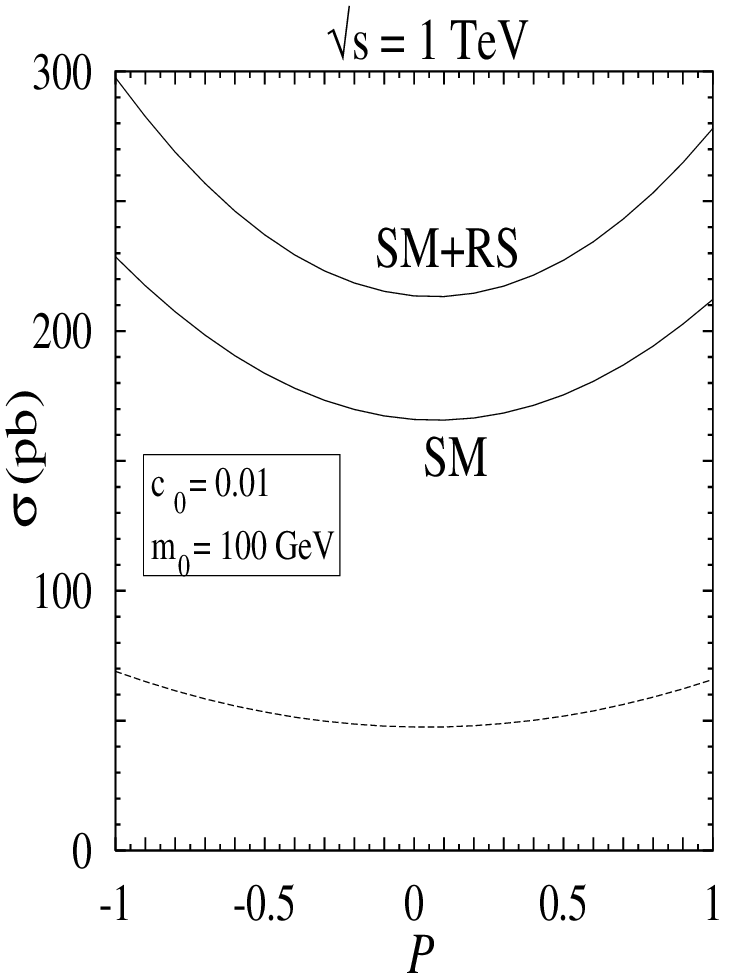}}
\end{center}
\caption{\footnotesize\sl
Variation of the cross-section for Moller scattering with ($a$) machine
energy and ($b$) polarization of the electron beams. In ($a$) the solid
curves correspond to the RS Model predictions for $m_0 = 150, 250$ and
500 GeV, while the dashed line represents the SM contribution. In ($b$),
the solid lines represent the SM and RS model contributions, while the
dashed line represents their difference. Other parameters are marked
(in the boxes).}
\end{figure}

In Fig. 2($b$), we present the variation of the cross-section with the
polarization $P$, at a 1~TeV machine, for the parameter set marked in the
inset box.  The solid curves correspond to the SM and the RS model
predictions, for a fixed set of parameters $(c_0,m_0)$, while the dashed
line represents the difference between the two. It is obvious that there
is a modest advantage to be gained from polarizing the beams, and there is
little difference between the cases when the beam is dominantly left- or
right-handed. This is also expected, since graviton exchanges are
non-chiral; in fact, the small difference arises from the interference
between diagrams with graviton and $Z$-exchange.

In order to estimate the discovery reach of a linear collider, we adopt
the following strategy. Discovery limits will be reached if the total
experimental cross-section agrees --- within the experimental precision
--- with the SM. Any excess or deficit must be attributed to BSM physics.
Thus, for a given energy $\sqrt{s}$, a given polarization $P$ and a fixed
set of parameters $(c_0,m_0)$, we calculate the total cross-section in
the RS model. A corresponding calculation of the SM cross-section,
multiplied by the luminosity, would lead to a predicted number of events.
We then estimate the errors assuming that the statistical errors are
Gaussian and that there are no systematic errors. While this certainly
makes our estimates of the discovery limits over-optimistic, we can argue
that electron detection efficiencies are generally high enough to allow us
to make a reasonable estimate in this approximation. In any case, before
more detailed studies of the detector design and systematic effects are
undertaken, any estimate of systematic errors must be pure guesswork.  We
choose, therefore, to neglect such effects. Finally the search reach of the
collider is given in terms of $3\sigma$ discovery limits.

\begin{figure}[h]
\begin{center}
\vspace*{3.35in}
      \relax\noindent\hskip -3.5in\relax{\includegraphics{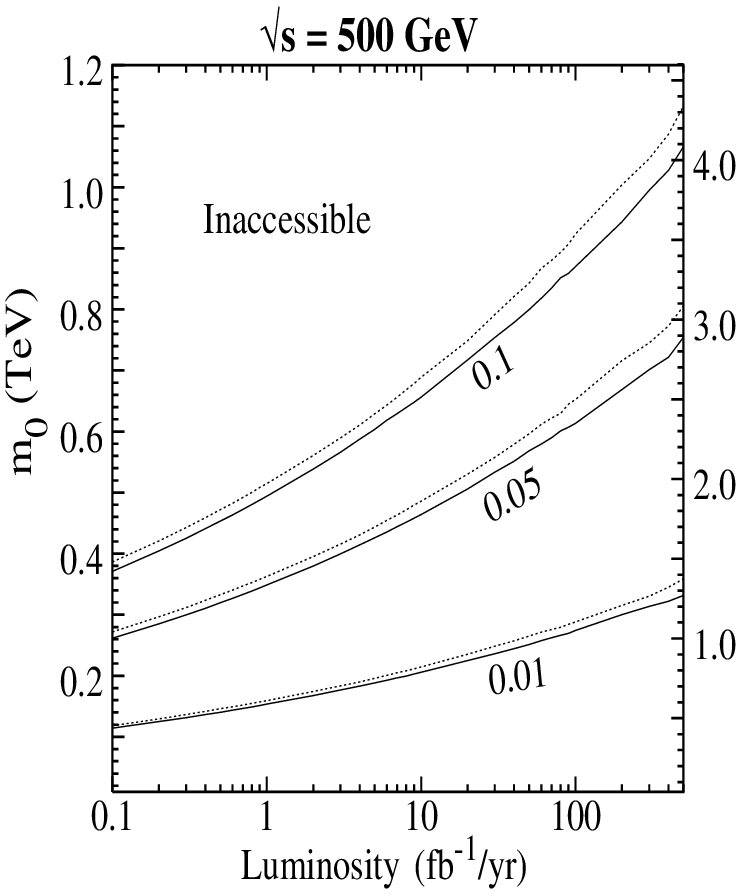}}
      \relax\noindent\hskip  3.4in\relax{\includegraphics{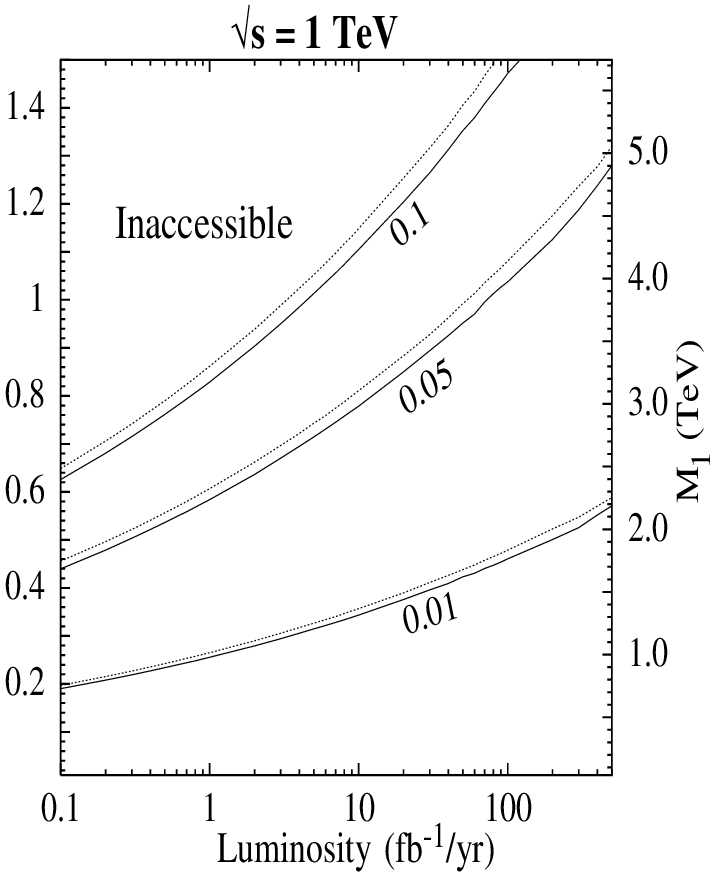}}
\end{center}
\caption{\footnotesize\sl
Discovery limits as a function of the integrated luminosity for
Randall-Sundrum graviton modes at an $e^- e^-$ collider at centre-of-mass
energies of 500 GeV and 1 TeV. Solid (dotted) lines correspond to
unpolarized (80\% left-polarized) electron beams.  The value of $c_0$ is
written alongside the relevant curve. The ordinate is labelled as a
function of the mass scale $m_0$ on the left and the mass of the lightest
resonance $M_1$ on the right.}
\end{figure}

Fig.~3 shows the search reach for the RS model at linear colliders running
at 500 GeV and 1 TeV respectively, as a function of the integrated
luminosity, for three different values of the coupling constant $c_0$
(marked along the curves). It may be seen that a linear collider could
easily probe $m_0$ up to at least 300 GeV --- which corresponds to a
lightest graviton mass of around 1.3 TeV --- if 500 pb$^{-1}$ of data are
collected. A slight improvement is possible with polarized beams, as the
dotted lines show. If the energy of the collider be increased to 1 TeV,
the reach goes up almost by a factor of 2. It may, then, be possible to
discover or exclude graviton resonances of mass 2.2 TeV or more.

While a 500 GeV or a 1 TeV collider will almost certainly be built, there
has been much interest in having a collider which probes the high energy
frontier\cite{Elis-Keil-Rldi}. In particular, it is possible that the CLIC
machine at CERN will be able to achieve a centre-of-mass energy as high as
3 TeV. Moreover, the possibility of a muon collider operating at a
centre-of-mass energy of 3--4 TeV has also received serious consideration.
For these machines, luminosities as high as $10^3$ fb$^{-1}$ per year have
also been considered. Gravitational effects in $e^- e^- \to e^- e^-$ are,
of course, identical to those in $\mu^- \mu^- \to \mu^- \mu^-$. In view of
these possibilities, we have explored the discovery reach of a 3 TeV
machine for the RS model. Our results are exhibited in Fig. 4. It may be
seen that this can easily probe $m_0$ as high as 1 TeV, which corresponds
to gravitons of mass nearly 4 TeV or more.

\begin{figure}[h]
\begin{center}
\vspace*{3.2in}
      \relax\noindent\hskip -5.0in\relax{\includegraphics{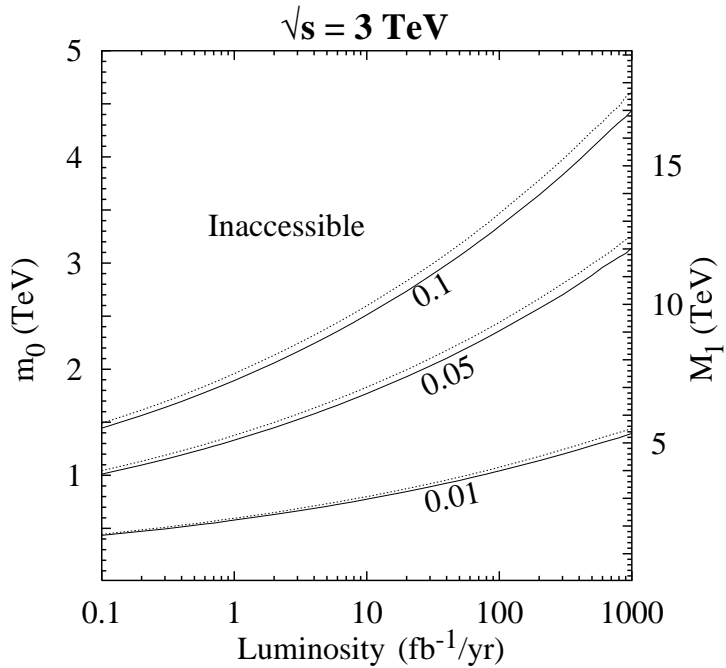}}
\end{center}
\caption{\footnotesize\sl
Discovery limits as a function of the integrated luminosity for
Randall-Sundrum graviton modes at an $e^- e^-$ collider at a
centre-of-mass energy of 3 TeV. The labelling is the same as in Fig.~2.  }
\end{figure}

It is worth noting that if graviton masses are pushed up to 5 TeV or more,
then, given that the scale ${\cal K}$ must be roughly an order of
magnitude smaller than $M_P^{(4)}$, it follows that the warp factor 
$e^{-\warpp}$ must be somewhat larger than is possible now. This would
either push up the Higgs boson mass to unacceptable values, or require
some mechanism to have a smaller mass scale origin for the Higgs boson
mass on the Planck brane. This would be a somewhat uncomfortable situation
for the RS model, since the original simplicity --- and therefore elegance
--- will be lost.

Finally we comment on the possibility of observing/constraining graviton
effects using the angular distribution of the final state electrons. Since
this form of BSM physics involves exchange of spin-2 particles, rather
than spin-1 particles, as in the SM, one can, in principle, expect a
rather different angular distribution 
for the electrons in the final state. In order to test this prediction, we
have made a $\chi^2$-analysis of the electron angular distribution in the
cases when there is graviton exchange and when there is no graviton
exchange. It turns out that the difference in the distributions is rather
small and confined to the central region. We find that one cannot get
better discovery limits by considering the angular distributions than
those which can be obtained by simply considering the total cross-section.
If indeed an excess or deficit over the SM prediction is found, angular
distributions might then become useful in determining the type of BSM
physics responsible, {\it e.g.} in distinguishing between spin-1 and
spin-0 exchanges. However, this would require high statistics and fine
resolutions. Accordingly, in this preliminary study, we do not pursue the
question of angular distributions any further.

In conclusion, therefore, an $e^- e^-$ collider would be a useful
laboratory to look for graviton exchange mechanisms, since there are very
few competing BSM processes. We find that a simple study of the total
cross-section for $e^- e^- \to e^- e^-$, subject to some minimal
acceptance cuts, leads to a prediction of rather optimistic discovery
limits. It is more useful to consider the total cross-section than the
angular distribution, which is rather similar to that in the SM.
Polarization of the beams can improve the search reach by a few percent,
irrespective of whether the beams are left- or right-polarized. At
a high energy collider, running at 3 or 4 TeV, the search limits can be
taken as far as graviton masses of 5 TeV or more, which is more-or-less
the frontier as far as the simplest version of the RS Model is concerned.

\bigskip\bigskip

\noindent
\footnotesize
{\bf Acknowledgements:} The authors would like to acknowledge Prasanta Das
and Saswati Sarkar for useful discussions, and the Theory Division, CERN for
hospitality while this work was being done. DKG would also like to thank
F.~Boudjema and LAPP, Annecy for hospitality.

\footnotesize
\newpage
\def\pr#1,#2,#3 { {\em Phys.~Rev.}        ~{\bf #1},  #2 (#3) }
\def\prd#1,#2,#3{ {\em Phys.~Rev.}        ~{\bf D#1}, #2 (#3) }
\def\prl#1,#2,#3{ {\em Phys.~Rev.~Lett.}  ~{\bf #1},  #2 (#3) }
\def\plb#1,#2,#3{ {\em Phys.~Lett.}       ~{\bf B#1}, #2 (#3) }
\def\npb#1,#2,#3{ {\em Nucl.~Phys.}       ~{\bf B#1}, #2 (#3) }
\def\prp#1,#2,#3{ {\em Phys.~Rept.}       ~{\bf #1},  #2 (#3) }
\def\zpc#1,#2,#3{ {\em Z.~Phys.}          ~{\bf C#1}, #2 (#3) }
\def\epj#1,#2,#3{ {\em Eur.~Phys.~J.}     ~{\bf C#1}, #2 (#3) }
\def\mpl#1,#2,#3{ {\em Mod.~Phys.~Lett.}  ~{\bf A#1}, #2 (#3) }
\def\ijmp#1,#2,#3{{\em Int.~J.~Mod.~Phys.}~{\bf A#1}, #2 (#3) }
\def\ptp#1,#2,#3{ {\em Prog.~Theor.~Phys.}~{\bf #1},  #2 (#3) }

\end{document}